\documentclass[twocolumn,showpacs,aps,prl,superscriptaddress]{revtex4}
\usepackage[dvips]{graphicx}
\usepackage{dcolumn}
\usepackage{bm}
\usepackage{multirow}
\begin{document}

\title{Cavity mode waves during terahertz radiation from rectangular Bi$_2$Sr$_2$CaCu$_2$O$_{8+\delta}$ mesas}
\author{Richard A. Klemm}\email{klemm@physics.ucf.edu}
\author{Erica R. LaBerge}
\author{Dustin R. Morley}
\affiliation{Department of Physics, University of Central Florida, Orlando, FL 32816, USA}
\author{Takanari Kashiwagi}
\author{Manabu Tsujimoto}
\author{Kazuo Kadowaki}\email{kadowaki@ims.tsukuba.ac.jp}
\affiliation{Graduate School of Pure \& Applied Sciences, University of Tsukuba, 1-1-1, Tennodai, Tsukuba, Ibaraki 305-8573, Japan}
\date{\today}
\begin{abstract}
We re-examined the angular dependence of the radiation from the intrinsic Josephson junctions in rectangular mesas of Bi$_2$Sr$_2$CaCu$_2$O$_{8+\delta}$, in order to determine if the cavity mode part of the radiation arises from waves across the width $w$ or along the length $\ell$ of the mesas, associated with ``hot spots'' [Wang {\it et al.}, Phys. Rev. Lett. {\bf 105}, 057002 (2010)].  We derived analytic forms for the angular dependence expected in both cases for a general cavity mode in which the width  of the mesa corresponds to an integer multiple of one-half the wavelength of the radiation.  Assuming the coherent radiation from the $ac$ Josephson current source and the cavity magnetic surface current density source combine incoherently, fits to the data of Kadowaki {\it et al.} [J. Phys. Soc. Jpn. {\bf 79}, 023703 (2010)] on a mesa with mean $\ell/w=5.17$  for both  wave directions using two models for the incoherent combination were made, which correspond to standing and traveling waves, respectively. The results suggest that the combined output from the uniform $ac$ Josephson current source plus a cavity wave forming along the rectangle length is equally probable as that of the combined output from the uniform $ac$ Josephson current plus a cavity wave across the width.  However, for mesas in which  $n\ell/2w$ is integral, where $n$ is the index of the rectangular TM$^z_{n,0}$ mode, it is shown that standing cavity mode waves  along the length of the mesa do not radiate in the $xz$ plane perpendicular to the length of the mesa, suggesting experiments on such mesas could help to resolve  the question.
\end{abstract}
\maketitle

\section{I. Introduction}
  Since there is a gap in the range 0.1-10 terahertz (THz) in the electromagnetic spectrum  for practical coherent sources, development of a material as a useful source could lead to many practical applications in the detection of complex molecules, in national security detection issues, space communications, environmental engineering, biochemistry analysis, cancer detection, and possibly genetic engineering \cite{Tonouchi}. Recently, the discovery of intense, continuous, coherent sub-terahertz radiation upon the application of a static voltage $V$ across the $N$ intrinsic Josephson junctions in  mesas of Bi$_2$Sr$_2$CaCu$_2$O$_{8+\delta}$ (Bi-2212) has generated a lot of excitement, in the hope that such a material may lead to a useful source filling the ``terahertz gap'' \cite{Ozyuzer,Kadowaki}.  As in those papers, it was generally found that the fundamental angular frequency $\omega_J$ of the $ac$ Josephson current generated by the applied $dc$ voltage $V$ satisfies the $ac$ Josephson relation
  \begin{eqnarray}
  \omega_J&=&
  \frac{2eV}{\hbar N},
  \end{eqnarray}
  where $e$ is the magnitude of the electronic charge, $\hbar$ is Planck's constant divided by $2\pi$, and the fundamental Josephson frequency $f_J=\omega_J/(2\pi)$.  An additional constraint on the radiation occurs when $V$ is adjusted in the mesa's current-voltage ($I/V$) characteristic such that the frequency matches that of an electromagnetic cavity mode of the rectangular mesa.
We assume the rectangular mesa has width $w$, length $\ell\ge w$, and height $h\ll w,\ell$.  Writing ${\bm k}$ and ${\bm x}$ for the wave vectors and position vectors outside the mesa, and ${\bm k}'$ and ${\bm x}'$ for the corresponding vectors inside the mesa, the angular frequencies $\omega_{mp}$ of the very thin rectangular cavity TM$^z_{m,p}$ modes are
\begin{eqnarray}
\omega_{mp}&=&ck_{mp}=\frac{ck_{mp}'}{n_r}=\frac{c\pi}{n_r}\Bigl[\Bigl(\frac{m}{w}\Bigr)^2+\Bigl(\frac{p}{\ell}\Bigr)^2\Bigr]^{1/2},
\end{eqnarray}
where $c$ is the speed of light in vacuum and $n_r=\sqrt{\epsilon}\approx4.2$ is the index of refraction of Bi-2212 in the frequency range of study\cite{Ozyuzer,Kadowaki}.  Hence, this additional constraint may be written as
\begin{eqnarray}
\omega_J&=&\omega_{m_0p_0},
\end{eqnarray}
where $m_0$ and $p_0$ are particular values of $m$ and $p$.
From fits to many rectangular mesas \cite{Kadowaki2}, it is evident that in all studies to date, the frequency $f_{1,0}$ of the TM$^z_{10}$ mode with $m_0=1,p_0=0$ matches the fundamental frequency $f_J$ of the radiation.  Therefore, most workers assumed that the fundamental mode was described by a standing wave of wavelength $\lambda=2w$ across the width of the mesa \cite{Ozyuzer,Kadowaki,Kadowaki2,BK1,BK2,LH1,LH2,LH3,HL1,HL2,MKM,MKMT,MHT,TFK,KMMK,KK,KK2}.  Many theoretical studies were made based upon the coupling of the ac Josephson current to this particular half-wavelength electromagnetic cavity mode \cite{BK1,BK2,LH1,LH2,LH3,HL1,HL2,MKM,MKMT,MHT,TFK,KMMK}.  However, with one exception \cite{KMMK}, these calculations were unable to predict any angular dependence of the output radiation from the cavity.  In the sole exception \cite{KMMK}, the mesa was modeled as a two-dimensional sheet in either the $xy$ plane or the $xz$ plane, and the output power was calculated numerically.  In that numerical calculation, it was therefore difficult to obtain the full three-dimensional angular dependence of the output power for comparison with experiment.  Thus, several workers employed the magnetic Love equivalence principle \cite{Love}, a variation on  Huygens' principle, to replace the electric field in the cavity by a magnetic surface current density, and then to use that effective magnetic surface current density to obtain the three-dimensional far-field radiation distribution\cite{HL2,KK,KK2}.

However, in fits to experiment on rectangular mesas, it was found that in addition to the radiation from the excitation of the electromagnetic cavity mode corresponding to the half-wavelength mode by the non-uniform part of the ac Josephson current, the uniform part of the ac Josephson current was also found to radiate\cite{Kadowaki2}.  This radiation was treated by the electric Love equivalence principle \cite{Love}, in which the ac Josephson current inside the mesa was effectively replaced by an electric surface current density\cite{KK,KK2}.  In fits to the observed output radiation from a rectangular mesa, it was then found that the observed radiation was consistent with a comparable amount from both the uniform ac Josephson current source and the electromagnetic cavity mode excited by the non-uniform part of the ac Josephson current in the mesa\cite{Kadowaki2}. Direct evidence for the importance of the uniform ac Josephson current source was subsequently provided by the investigation of cylindrical mesas\cite{Kadowaki3}.  In that experiment, three disk-shaped cylindrical cavities were found to exhibit radiation not only at the fundamental electromagnetic cavity mode for a cylindrical cavity, but also at the integral second harmonic of the fundamental radiation frequency \cite{Kadowaki3}. Since the modes for a cylindrical cavity are not integral multiples of one another\cite{KK2,HL2}, it was therefore determined that the observed output radiation was a combination of that arising from the excitation of the fundamental cylindrical cavity mode excited by the main non-uniform part of the ac Josephson current and the radiation from the uniform part of the ac Josephson current itself \cite{KK2,Kadowaki3}.  The integrated intensities of the outputs from those two sources were found to be comparable to one another\cite{Kadowaki3}.  Hence, the dual-source mechanism of radiation from both the uniform part of the ac Josephson current and from the excitation of a cavity mode from the appropriate part of the non-uniform ac Josephson current was established\cite{Kadowaki3}.

However, the observation of ``hot spots'' in rectangular Bi-2212 mesas accompanied by the apparent evidence of standing waves with $\lambda\approx 2w$ along the length of the mesa have led to a lot of interest \cite{Kleiner}.  Very recently, Wang {\it et al.} provided evidence that the samples exhibiting such hot spots do in fact emit radiation \cite{Kleiner2}. Although there is widespread agreement that the fundamental mode in all rectangular mesas studied to date has the frequency of the TM$^z_{1,0}$ mode, it has therefore been an open question as to whether the waves contributing to the radiation form across the width of the mesa, as expected from standard cavity theory \cite{antenna,antenna2}, or form with the same wavelength $\lambda=2w$ along the length of the mesa.  In this paper, we address that issue in detail, performing detailed fits to the data of Kadowaki {\it et al.} for the angular dependence of the radiation in the $xz$ and $yz$ planes normal and parallel to the mesa length, respectively \cite{Kadowaki2}, and find that both cavity wave directions make equally probable contributions to the radiation for that mesa with a mean $\ell/w=5.17$.  However, we show that in the special cases in which $n\ell/w=2q$, where $q$ is a natural number and $n$ is the index of the TM$^z_{n,0}$ mode, an excited cavity wave formed along the length of the mesa cannot emit radiation in the $xz$ plane normal to the mesa length, and we thus propose detailed angular dependence studies of the radiation from such mesas as a more precise test of this question.

In Sec. II, we present the angular dependence of the radiation from a rectangular mesa with the radiation arising from two sources:  the uniform $ac$ Josephson current source, and the radiation from an excited electromagnetic cavity mode, in which either standing or traveling waves may form either across the width or along the length of the mesa.  In Sec. III, we present our fits to the available published data on the angular dependence of the radiation from rectangular cavities.  In Sec. IV, we consider the special cases of  $n\ell/w=2q$.  Finally, in Sec. V, we summarize and discuss our results.

\section{II. Angular dependence of the radiation}

From previous studies, it was determined that the radiation from rectangular and cylindrical mesas of Bi-2212 arose from a dual-source mechanism \cite{Kadowaki2,Kadowaki3}, in which the magnetic field associated with the uniform part of the $ac$ Josephson current  and the electric field associated with the $n$th $ac$ Josephson current harmonic at $f_n=nf_J$ and the $n$th cavity mode of the mesa were treated using the Love equivalence principles as electric and magnetic surface current sources ${\bm J}_{Sn}({\bm x}',t)$ and ${\bm M}_{Sn}({\bm x}',t)$, respectively \cite{Kadowaki2,KK,KK2,antenna,antenna2,Kadowaki3,Love}.
In rectangular source coordinates $(x',y',z')$ for the directions along the width, length, and height of the mesa, respectively, for an electric surface current density ${\bm J}_{Sn}({\bm x}')$ and either a magnetic surface current density ${\bm M}_{Sn,w}({\bm x}')$ or ${\bm M}_{Sn,\ell}({\bm x}')$ for the $n$th mode forming across the width or along the length of the mesa, respectively,
\begin{eqnarray}
{\bm J}_{Sn}({\bm x}')&=&\hat{\bm z}'\frac{J_n^J}{4}\eta_J(z')\sum_{\sigma=\pm}[f_{\sigma}(x',y')+g_{\sigma}(x',y')],\nonumber\\
& &\label{Jrect}\\
{\bm M}_{Sn,w}({\bm x}')&=&\frac{\tilde{E}_{0n0}}{4}\eta_M(z')\sin[n(x'-x_n)\pi/w]\nonumber\\
& &\times\sum_{\sigma=\pm}\sigma[\hat{\bm y}'f_{\sigma}(x',y')-\hat{\bm x}'g_{\sigma}(x',y')],\label{Mwidth}\\
{\bm M}_{Sn,\ell}({\bm x}')&=&\frac{\tilde{E}_{0n0}}{4}\eta_M(z')\sin[n(y'-y_n)\pi/w]\nonumber\\
& &\times\sum_{\sigma=\pm}\sigma[\hat{\bm y}'f_{\sigma}(x',y')-\hat{\bm x}'g_{\sigma}(x',y')],\label{Mlength}\\
f_{\sigma}(x',y')&=&w\delta(x'+\sigma w/2)\Theta[(\ell/2)^2-(y')^2],\label{f}\\
g_{\sigma}(x',y')&=&{\ell}\delta(y'+\sigma \ell/2)\Theta[(w/2)^2-(x')^2],\label{g}
\end{eqnarray}
where $\tilde{E}_{0n0}$ and $J_n^J$ are the amplitudes of the electric field for the TM$^z_{n,0}$ mode and the $n$th harmonic of the $ac$ Josephson current at the surface, and the TM$^z_{n,0}$ cavity mode energy is degenerate for either $-w/n\le x_n\le w/n$ or for $-w/n\le y_n\le w/n$.  The substrate factors $\eta_J(z')$ and $\eta_M(z')$ may be written as $\Theta(z')\Theta(h-z')$ for a sample suspended in vacuum, but for a sample on a superconducting substrate, they may be approximated as ${\rm sgn}(z')\Theta[h^2-(z')^2]$\cite{KK,KK2}. Using the Schelkunoff procedure of adding the contributions from these sources to the magnetic and electric vector potentials ${\bm A}({\bm x},t)$, ${\bm F}_w({\bm x},t)$ and/or ${\bm F}_{\ell}({\bm x},t)$ in the radiation zone far from  the mesa \cite{Schelkunoff,antenna,antenna2},  we have
\begin{eqnarray}
{\bm A}({\bm x},t)&=&\frac{\mu_0}{8\pi}\sum_{n=1}^{\infty}\int d^3{\bm x}'{\bm J}_{Sn}({\bm x}')\frac{e^{in(k_JR-\omega_Jt)}}{R},\label{A}\\
{\bm F}_i({\bm x},t)&=&\frac{\epsilon_0}{8\pi}\sum_{n=1}^{\infty}\int d^3{\bm x}'{\bm M}_{Sn,i}({\bm x}')\frac{e^{in(k_JR-\omega_Jt)}}{R},\label{Fi}
\end{eqnarray}
where $i=w,\ell$, respectively, $R=|{\bm x}-{\bm x}'|$,
and we use the well-established far-field approximation
\begin{eqnarray}
\frac{e^{ikR}}{R}&\rightarrow&\frac{e^{ikr}}{r}e^{-i{\bm k}\cdot{\bm x}'},
\end{eqnarray}
where $r=|{\bm x}|$.
With this approximation, the integrals in Eqs. (\ref{A}) and (\ref{Fi}) are elementary, and
 ${\bm A}({\bm x},t)$ and the ${\bm F}_i({\bm x},t)$ in spherical coordinates are  given in the radiation zone by
\begin{eqnarray}
{\bm A}({\bm x},t)&\rightarrow&\frac{\mu_0\hat{\bm z}\tilde{v}}{8\pi r}\sum_{n=1}^{\infty}J_n^Je^{in(k_Jr-\omega_Jt)}S^J_n(\theta)\chi_n,\label{Anrect}\\
\chi_n(\theta,\phi)&=&\cos X_n\frac{\sin Y_n}{Y_n}+\cos Y_n\frac{\sin X_n}{X_n},\label{chin}\\
{\bm F}_i({\bm x},t)&\rightarrow&-\frac{\epsilon_0\tilde{v}}{16\pi r}\sum_{n=1}^{\infty}\tilde{E}_{0n0}e^{in(k_Jr-\omega_Jt)}\nonumber\\
& &\times S^M_n(\theta)
(\hat{\bm x}M^x_{n,i}+\hat{\bm y}M^y_{n,i}),\\
M_{n,w}^x&=&-\sin Y_n\sum_{\sigma=\pm}\sigma e^{i\sigma n\pi x_n/w}\frac{\sin X_{n,\sigma}}{X_{n,\sigma}},\label{Mnwx}\\
M_{n,w}^y&=&\frac{\sin Y_n}{Y_n}\sum_{\sigma=\pm}e^{i\sigma X_n}\sin\Bigl(\frac{n\pi}{2}+\frac{\sigma n\pi x_n}{w}\Bigr),\label{Mnwy}\\
M_{n,\ell}^x&=&\frac{\sin X_n}{X_n}\sum_{\sigma=\pm}e^{i\sigma Y_n}\sin\Bigl(\frac{n\pi\ell}{2w}+\frac{\sigma n\pi y_n}{w}\Bigr),\label{Mnlx}\\
 M_{n,\ell}^y&=&-\sin X_n\sum_{\sigma=\pm}\sigma e^{i\sigma n\pi y_n/w}\frac{\sin Y_{n,\sigma}}{Y_{n,\sigma}},\label{Mnly}
\end{eqnarray}
where $X_n=(k_nw/2)\sin\theta\cos\phi$,
$Y_n=(k_n{\ell}/2)\sin\theta\sin\phi$, $k_nw=n\pi/n_r$,  $X_{n,\sigma}=n\pi/2+\sigma X_n$, $ Y_{n,\sigma}=\frac{n\pi\ell}{2w}+\sigma Y_n$, $\tilde{v}=w{\ell}h$, $S^M_n(\theta)=S^J_n(\theta)=1$ for no substrate,  $\hat{\bm x}=\hat{\bm r}\sin\theta\cos\phi+\hat{\bm\theta}\cos\theta\cos\phi-\hat{\bm\phi}\sin\phi$, $\hat{\bm y}=\hat{\bm r}\sin\theta\sin\phi+\hat{\bm\theta}\cos\theta\sin\phi+\hat{\bm\phi}\cos\phi$, and $x_n, y_n$ appear in Eqs. (\ref{Mwidth}) and (\ref{Mlength}), respectively.

Using the Schelkunoff procedure that ${\bm E}_{\bm A}=-\frac{\partial {\bm A}}{\partial t}$ and ${\bm H}_{\bm A}=\frac{1}{\mu_0}{\bm \nabla}\times{\bm A}$, ${\bm E}_{\bm F}=-\frac{1}{\epsilon_0}{\bm \nabla}\times{\bm F}$, and $-\frac{\partial{\bm H}_{\bm F}}{\partial t}={\bm\nabla}\times{\bm E}_{\bm F}$, we write ${\bm E}={\bm E}_{\bm A}+{\bm E}_{\bm F}$ and ${\bm H}={\bm H}_{\bm A}+{\bm H}_{\bm F}$. Then, the differential power per unit solid angle in the radiation zone is given by $dP/d\Omega=\frac{1}{2}{\rm Re}[r^2\hat{\bm r}\cdot\overline{{\bm E}\times{\bm H}^{*}}]$, where the overline denotes a time average \cite{KK2,antenna,antenna2,Schelkunoff}.
Because of the nearly symmetric radiation observed on both sides and on both ends of the mesa, with maximal radiation in the $xz$ plane with $\theta\approx\pm30^{\circ}$ \cite{Kadowaki2},  we study the incoherent mixture of the coherent radiation from both the uniform $ac$ Josephson current and the electromagnetic cavity sources in two models.  In Model I, we preserve the von Neumann boundary condition that the tangential component of the magnetic field vanishes at the boundary, which becomes either $H_y(x'=\pm w/2)=0$ for standing waves across the width or $H_x(y'=\pm\ell/2)=0$ for standing waves  centered symmetrically or antisymmetrically, depending upon $\ell/w$, along the length.  This  leads to the respective solutions $x_n=0,w/n$ (or $y_n=0,w/n$) for $n$ odd, and $x_n=\pm w/2n$ (or $y_n=\pm w/2n$) for $n$ even, and we average $dP/d\Omega$ over these two solutions.  In Model II, we relax the boundary condition, and average $dP/d\Omega$ over the allowed range of values (from $-w/n$ to $w/n$) of  either  $x_n$ or $y_n$.  This model corresponds to traveling waves, as the positions of the wave nodes during the time of the measurement can be considered to have moved an arbitrary amount.
Then, it is straightforward to obtain  the time-averaged $dP_{i}^j/d\Omega$  in the radiation zone \cite{KK,KK2},
 \begin{eqnarray}
 \frac{dP^j_i}{d\Omega}&=&\sum_{n=1}^{\infty}\frac{dP^j_{n,i}}{d\Omega},\\
 \frac{dP^j_{n,i}}{d\Omega}&{{\rightarrow}\atop{r/a\rightarrow\infty}}&\frac{Z_0(\tilde{v}k_J)^2}{128\pi^2}n^2\biggl[\Bigl|\sin\theta J^J_n\chi_n(\theta,\phi)S^J_n(\theta)\Bigr|^2\nonumber\\
 & &\>+\alpha_n(\theta)\Bigl(C^j_{n,i}+D^j_{n,i}-\sin^2\theta[C^j_{n,i}\cos^2\phi\nonumber\\
 & &+D^j_{n,i}\sin^2\phi-E^j_{n,i}\sin\phi\cos\phi]\Bigr)\biggr],\label{Prect}
 \end{eqnarray}
 where $j =$ I, II, $i = w, \ell$, $\alpha_n(\theta)=|\tilde{E}_{0n0}S^M_n(\theta)|^2/(2Z_0)^2$,  $Z_0$ is the vacuum impedance, $k_J=c\omega_J$, and the expressions for the $C_{n,i}^j$, $D_{n,i}^j$, and $E_{n,i}^j$ are given in the Appendix.  For a superconducting substrate, we take  $S^M_n(\theta)\propto S^J_n(\theta)\propto \cos\theta$ \cite{KK,KK2}.

  The  expressions listed in the Appendix exhibit the symmetries that the pairs $D_{n,i}^{ j}$ and $C_{n,i'}^{j}$ and the pairs $E_{n,i}^j$ and $E_{n,i'}^j$ for $i\ne i'$ are related by interchanging $X_n$ with $Y_n$ and $X_{n,\sigma}$ with $Y_{n,\sigma}$.    The relations between the Model I quantities are broken into even and odd values of $n$.  We note that for square samples, $\ell/w=1$, these symmetries merely reflect a rotation by $90^{\circ}$ about the $z$ axis.  In this case, radiation from either wave formation direction is equally allowed \cite{Kadowaki3}.
 \begin{table}
  \begin{tabular}{cclllll}
  \hline
\hline
\noalign{\vskip3pt}
  $i$&$j$&${\cal A}_{1,i}^j$&${\cal B}_{1,i}^j$&$\sigma_{1,i}^j$&$\overline{g_{1,i}^j}$&$\overline{P}_{\rm cav}/\overline{P}_{\rm tot}$\\
  $w$&I&0.4307&0.1335&0.1235&5.431&0.491\\
  $w$&II&0.4241&0.2592&0.1238&3.343&0.539\\
  $\ell$&I&0.5606&0.1436&0.1250&3.320&0.328\\
  $\ell$&II&0.5567&0.2873&0.1234&1.764&0.343\\
  \hline
\hline
  \end{tabular}
  \caption{Least-squares fitting parameters ${\cal A}_{1,i}^j$ and ${\cal B}_{1,i}^j$ for the output  $dP_{1,i}^j/d\Omega={\cal A}_{1,i}^jf_1(\theta,\phi)+{\cal B}_{1,i}^jg_{1,i}^j(\theta,\phi)$ and the standard deviations $\sigma_{1,i}^j$ obtained by assuming the waves form across the width and along the length in Models I and II.  The integrated power $\overline{P}_{\rm tot}={\cal A}_{1,i}^j\overline{f_1}+{\cal B}_{1,i}^j\overline{g_{1,i}^j}$, where $\overline{f_1}=1.744$ for $\ell/w=400/77.4$ and $n_r=4.2$.  The portion  of the integrated output power arising from the respective cavity mode  is $\overline{P}_{\rm cav}={\cal B}_{1,i}^j\overline{g_{1,i}^j}$. See text.}
  \end{table}
\begin{figure}
\includegraphics[width=0.9\columnwidth]{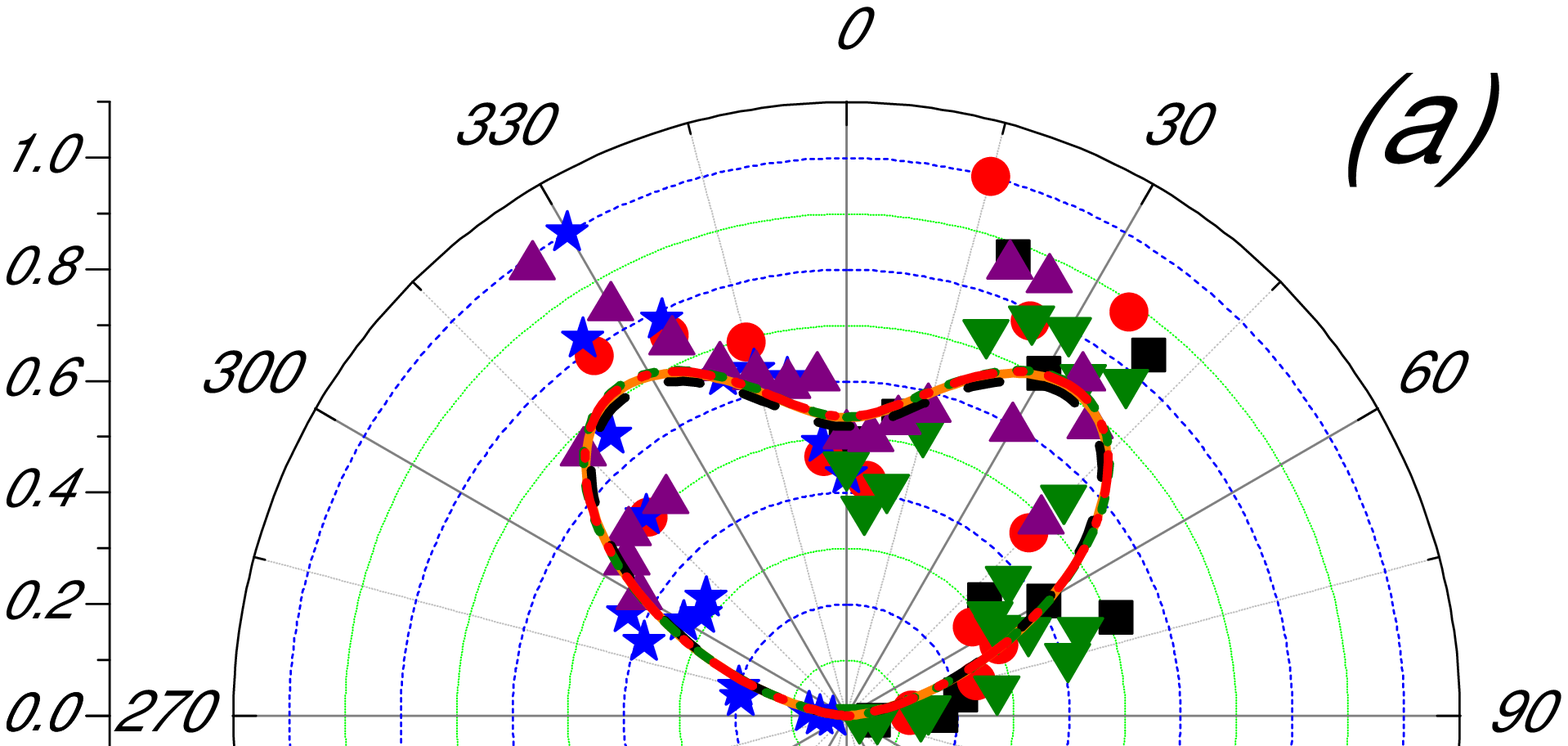}\hskip10pt
\includegraphics[width=0.9\columnwidth]{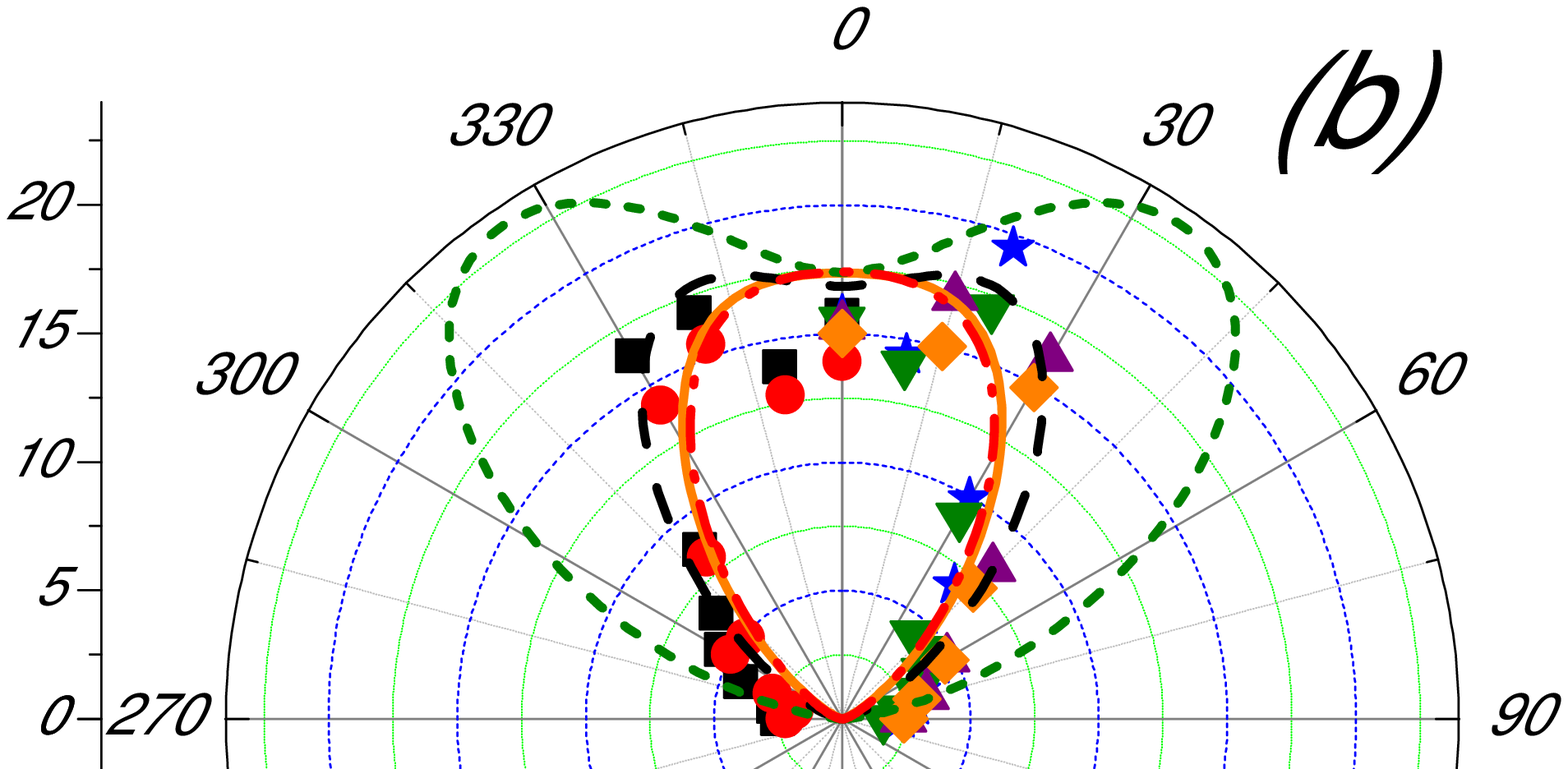}
\caption{(color online)   Polar ($\theta$ in degrees) plots of the four least-squares fits to the data taken in individual planes. (a) the $xz$ plane ($\phi=0^{\circ}$) normal to the mesa length.  (b) The $yz$  plane ($\phi=90^{\circ}$) normal to the mesa width.  Solid orange curves: $w$, I.  Dashed black curves:  $w$, II.  Short-dashed olive curves:  $\ell$, I.  Dash-dotted red curves:  $\ell$, II. Data are from Kadowaki {\it et al.}\cite{Kadowaki2}.  See text.}\label{fig1}
\end{figure}

\begin{figure}
\includegraphics[width=0.45\textwidth]{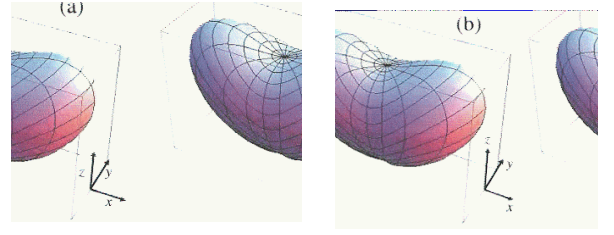}
\caption{(color online) (a) Three-dimensional plot of the best fit of $dP_{1,w}^{\rm I}(\theta,\phi)/d\Omega$ to the angular data of Kadowaki {\it et al.}\cite{Kadowaki2},   obtained with the standing wave across the width.  (b) Three-dimensional plot of the best fit of $dP_{1,\ell}^{\rm II}(\theta,\phi)/d\Omega$ to the same data obtained with the traveling wave across the length.}\label{fig2}
\end{figure}

 \section{III. Fits to the available data on a rectangular mesa}
  We have fit the angular dependence of the radiation from the rectangular mesa of Kadowaki {\it et al.} \cite{Kadowaki2}, using Models I and II for the $n=1$ mode, for both  waves forming across the width and along the length, including the uniform $ac$ Josephson current in all cases.  We note that the rectangular axes were defined by Kadowaki {\it et al.} to be in accordance with antenna theory \cite{antenna,antenna2}, with the $x, y,$ and $z$ axes respectively along the height, length, and width of mesa, so that $\theta$ was the angle of a vector from the $x$ axis and $\phi$ was the angle of the projection of the vector in the $yz$ plane from the $y$ axis \cite{Kadowaki2}.  In this work, we assume the standard definitions of spherical coordinates that the $x,y,z$ axes correspond respectively to the directions across the width, length, and height of the mesa, so that $\theta$ is the angle a vector makes with the $z$ axis, and $\phi$ is the angle the projection of the vector in the $xy$ plane makes with the $x$ axis. There are four values for the fitting parameters, which we denote ${\cal A}_{1,i}^j$  and ${\cal B}_{1,i}^j$, where $i = w, \ell$, and $j$ = I, II. These ${\cal A}_{1,i}^j$ and ${\cal B}_{1,i}^j$ are fitting parameters of the theoretical functions to the data,
 \begin{eqnarray}
 \frac{dP_{1,i}^j}{d\Omega}(\theta,\phi)&=&{\cal A}_{1,i}^jf_1(\theta,\phi)+{\cal B}_{1,i}^jg_{1,i}^j(\theta,\phi),\\
 f_n(\theta,\phi)&=&\sin^2\theta\cos^2\theta\chi_n^2(\theta,\phi),\\
 g_{n,i}^j(\theta,\phi)&=&\cos^2\theta[C_{n,i}^j+D_{n,i}^j-\sin^2\theta(C_{n,i}^j\cos^2\phi\nonumber\\
 & &+D_{n,i}^j\sin^2\phi-E_{n,i}^j\sin\phi\cos\phi)],
 \end{eqnarray}
  where the $\chi_n(\theta,\phi)$ are given by Eq. (\ref{chin}) and the  $C_{n,i}^j(\theta,\phi)$, $D_{n,i}^j(\theta,\phi)$ and  $E_{n,i}^j(\theta,\phi)$ are given in the Appendix.

  The $n=1$ mode of a rectangular  mesa with mean $\ell/w=400/77.4\approx5.17$ is the only sample for which the angle dependence of the radiation in both the $xz$ and $yz$ planes is available \cite{Kadowaki2}.  The least-squares fits include the 85 data points in the $xz$ plane $(\phi=0^{\circ})$ and the 54 data points in the $yz$ plane $(\phi=90^{\circ})$ with  $-90^{\circ}\le\theta\le90^{\circ}$.  The nearly symmetric distribution about $\theta=0^{\circ}$ implies a rather incoherent mixture of the two radiation sources \cite{Kadowaki2,KK,KK2}.  In Table I, a list of the $n=1$ fitting parameters ${\cal A}_{1,i}^j$ and ${\cal B}_{1,i}^j$ for $j$ = I, II and $i = w, \ell$ is given.  In addition, numerical values for the integrated components $\overline{g_{1,i}^j}$ to the output power $\overline{P_{1,i}^j}=\int_0^{\pi/2}\sin\theta d\theta\int_0^{2\pi}d\phi dP_{1,i}^j(\theta,\phi)/d\Omega$  from the respective cavity modes with $\ell/w=400/77.44$  and $n_r=4.2$ are given, and from the listed ${\cal A}_{1,i}^j$ and ${\cal B}_{1,i}^j$ values and   the analogous integral $\overline{f_1}=1.744$, the fraction $\overline{P}_{\rm cav}/\overline{P}_{\rm tot}=[1+{\cal A}_{1,i}^j\overline{f_1}/({\cal B}_{1,i}^j\overline{g_{1,i}^j})]^{-1}$ of the integrated output power arising from the respective cavity mode is also given.  We remark that the standard deviations of all of the fits are very similar, so that  either cavity wave formation direction is equally probable. However, the nearly identical overall best fits are for the standing wave across the width (Model I) and  the traveling wave along the length (Model II), as seen in Table I.

  In Fig. 1, polar ($\theta$) plots of the best two-parameter fits to the experimental data of Kadowaki {\it et al.} in the $xz$ and $yz$ planes for each of the four models are shown.  While all four models give nearly identically good fits to the data in the $xz$ plane, as shown in Fig. 1(a), the third  model with the wave across the length in Model I leads to a distinctly worse fit to the data in the $yz$ plane, as shown by the short-dashed olive curve in Fig. 1(b).  In addition, Model II for the traveling wave across the width does not provide quite as good a fit in the $yz$ plane as do Model I for the standing wave across the width and Model II for the traveling wave across the length.  Moreover, there are some physical differences in the nature of the fits due to the dual-source mechanism.  In the cases of an excited cavity mode which is either a standing or traveling wave across the width, nearly equal proportions of the radiation arise from the excited cavity mode and from the uniform $ac$ Josephson current source \cite{Kadowaki2}.  On the other hand, if the excited cavity mode were to result in either a standing or traveling wave  along the length of the mesa, only about one-third of the radiation would arise from the excited cavity mode, and about two-thirds of it would arise from the uniform $ac$ Josephson current source, as indicated in Table I.

  Three-dimensional plots of the predicted angular dependence of the output radiation obtained from the two best fits are shown in Fig. 2.  As in Kadowaki {\it et al.} \cite{Kadowaki2}, Fig. 2(a) is for the dual-source mechanism with the standing wave across the mesa width in Model I, preserving  the von Neumann boundary conditions.  The angular dependence of the output radiation obtained  for the traveling wave along the mesa length in Model II is shown in Fig. 2(b). In this case, the wave can move back and forth along the length without regard to the von Neumann boundary conditions.  As seen from Figs. 1 and 2, it is nearly impossible to decide visually which fit is better, as the three-dimensional patterns formed in the radiation zone are nearly identical, and the standard deviations only differ by $7\times10^{-5}$, which is completely negligible.

  \begin{figure}
\includegraphics[width=0.45\textwidth]{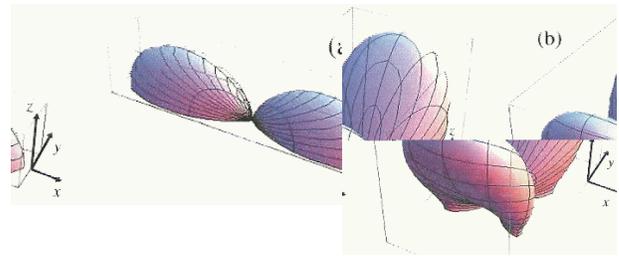}
\caption{(color online) Spherical plots of the electromagnetic cavity radiation from a rectangular mesa with $\ell/w=4$ and $n_r=4.2$  suspended in vacuum. (a) The TM$^z_{2,0}$ mode  (b) The TM$^z_{0,8}$ mode.}\label{fig3}
\end{figure}
\begin{figure}
\includegraphics[width=0.45\textwidth]{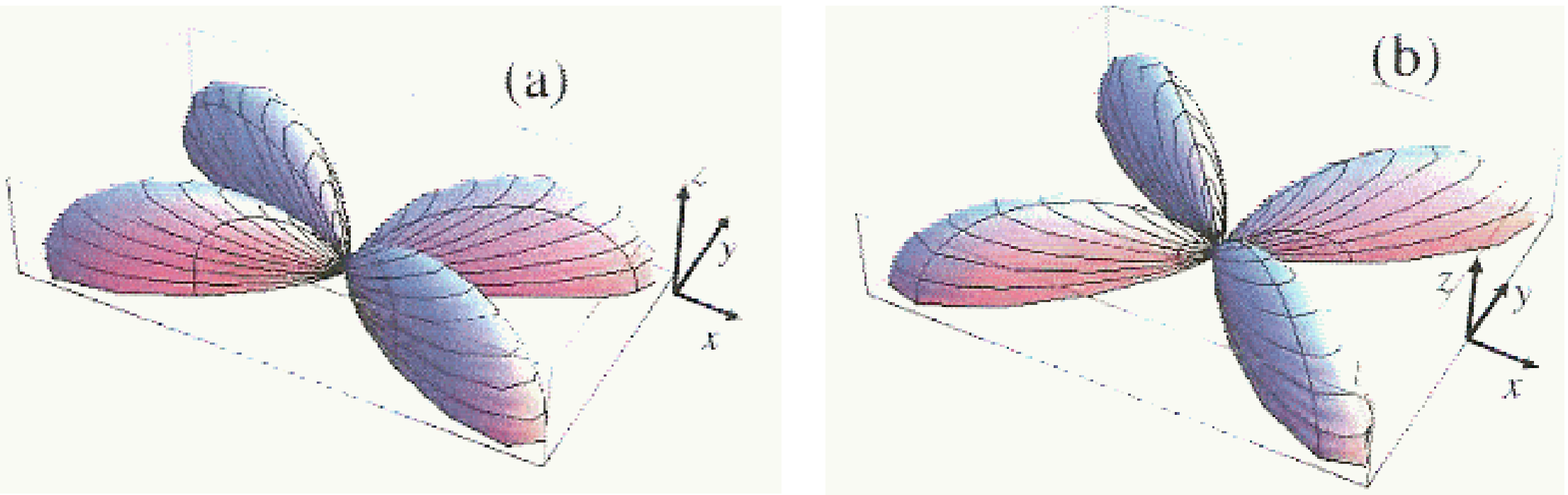}
\caption{(color online) Spherical plots of the electromagnetic cavity radiation from a rectangular mesa with $\ell/w=4$ and $n_r=4.2$  suspended in vacuum. (a) The TM$^z_{2,1}$ mode  (b) The TM$^z_{2,2}$ mode.}\label{fig4}
\end{figure}

\begin{figure}
\includegraphics[width=0.45\textwidth]{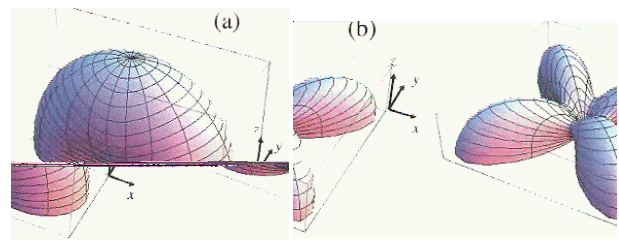}
\caption{(color online) Spherical plots of the electromagnetic cavity radiation from a square mesa with $\ell/w=1$ and $n_r=4.2$  suspended in vacuum. (a) The TM$^z_{1,0}$ mode  (b) The TM$^z_{1,1}$ mode.}\label{fig5}
\end{figure}
 \section{IV. Special cases}

 In order to guide future experiments to help determine which cavity mode excitation actually radiates, we now examine the radiation from the cavity source in some special situations.  We are particularly interested in the azimuthal angles $\phi=0, 90^{\circ}$, for which the radiation is in the $xz$ and $yz$ planes, respectively.  Let us consider the special case
 \begin{eqnarray}
 \frac{n\ell}{2w}&=&q,\label{integer}
 \end{eqnarray}
 where $q$ is any integer.   For this special case, it is easy to see from Eqs. (\ref{Mnlx}) and (\ref{Mnly}) that for $\phi=0^{\circ}$, $Y_n=0$, $Y_{n,\sigma}=q\pi$, and hence for arbitrary $y_n$, $M_{n,\ell}^x=M_{n,\ell}^y=0$. Thus, for either a pure cavity radiation source or a coherent or incoherent mixture of the two coherent radiation sources, for $n\ell/w=2q$, one would not expect to obtain any radiation in the $xz$ plane from an excited cavity mode leading to a wave along the length of the mesa.  Hence, for these special cases, the radiation in the $xz$ plane is expected to arise  from the uniform $ac$ Josephson current source and/or from the excited cavity mode forming a wave across the width.

 There are two different examples  of  the special case.  If $n=1$ or some odd integer, for the special case to be realized, we must have $\ell/w=2q$ in order that no radiation in the $xz$ plane from cavity modes forming  waves along the length could radiate.  In the case $n=2$ or any higher even value, this special case would apply to any integral value of $\ell/w$.

 On the other hand, symmetry suggests we also examine the $yz$ plane, for which $\phi=90^{\circ}$.  In this case with $n\ell/w=2q$ and $\phi=90^{\circ}$, $X_n=0$.  Although for $n$ odd, $X_{n,\sigma}=n\pi/2$ is an odd multiple of $\pi/2$, for even $n$, $X_{n,\sigma}=n\pi/2$ is an integral multiple of $\pi$.  As seen from Eqs. (\ref{Mnwx}) and (\ref{Mnwy}), $n$ even and $\phi=90^{\circ}$ implies $M_{n,w}^x=M_{n,w}^y=0$ for any value of $x_n$, and the radiation from the excited cavity mode forming a  wave across the width vanishes in the $yz$ plane.    Thus, examination of rectangular mesas with integral $\ell/w$ that radiate in the TM$^z_{2,0}$ mode would be especially illuminating to separate out the possible standing wave contributions to the radiation.

 We note that when $n\ell/w=p$ is an integer, $f_{n,0}=f_{0,p}$, so the energies of the modes with $n$ half-integral wavelengths across the width are degenerate with those of $p$ half-integral wavelengths along the length.  Thus, simultaneous radiation from both cavity modes is possible. The amplitudes of the electric field for these cases may be written as $\tilde{E}_{0n0}$ and $\tilde{E}_{00p}$, which are not necessarily
 identical.  In this case, the combined radiation at the $n$th mode frequency $f_{n,0}$ may be written as a modified version of Eq. (\ref{Prect}), as
\begin{eqnarray}
\frac{dP^j_{n}}{d\Omega}&{{\rightarrow}\atop{r/a\rightarrow\infty}}&\frac{Z_0(\tilde{v}k_J)^2}{128\pi^2}n^2\biggl[\Bigl|\sin\theta J^J_n\chi_n(\theta,\phi)S^J_n(\theta)\Bigr|^2\nonumber\\
 & &\>+\alpha_n(\theta)\Bigl(\tilde{C}^j_{n}+\tilde{D}^j_{n}-\sin^2\theta[\tilde{C}^j_{n}\cos^2\phi\nonumber\\
 & &+\tilde{D}^j_{n}\sin^2\phi-\tilde{E}^j_{n}\sin\phi\cos\phi]\Bigr)\biggr],
 \end{eqnarray}
 where
 \begin{eqnarray}
 \tilde{C}^j_{n}&=&C_{n,w}^j+R_nD_{p,\ell}^j,\\
 \tilde{D}^j_{n}&=&D_{n,w}^j+R_nD_{p,\ell}^j,\\
 \tilde{E}^j_{n}&=&E_{n,w}^j+R_nE_{p,\ell}^j,
 \end{eqnarray}
  where $R_n=|\tilde{E}_{00p}|^2/|\tilde{E}_{0n0}|^2$, and the $C_{m,i}^j$, $D_{m,i}^j$, and $E_{m,i}^j$ for $m=n,p$ are given in the Appendix by setting $k_n=k_{n0}=k_{0p}$.    Thus, when $n\ell/w=p$, in principle there will be three contributions to the radiation at a given frequency:  that due to the uniform $ac$ Josephson current source, and that due to the excitations of the degenerate cavity TM$^z_{n,0}$ and TM$^z_{0,p}$ modes.  Hence, a three parameter fit is required to determine the relative contribution of each source.

 We now consider the case that the rectangular cavity is excited into a non-degenerate TM$^z_{m,p}$ mode with $m,p\ne0$  by resonance with the fundamental $ac$ Josephson current angular frequency $\omega_J=\omega_{mp}=ck_{mp}$ given by Eqs. (2) and (3).   The magnetic surface current density for the TM$^z_{m,p}$ mode may be written as
\begin{eqnarray}
{\bm M}_{Smp}({\bm x}')&=&\frac{\tilde{E}_{0mp}}{4}\eta_M(z')\sin[m(x'-x_m)\pi/w]\nonumber\\
& &\times\sin[p(y'-y_p)\pi/\ell]\nonumber\\
& &\times\sum_{\sigma=\pm}\sigma[\hat{\bm y}'f_{\sigma}(x',y')-\hat{\bm x}'g_{\sigma}(x',y')],\label{Mrect}
\end{eqnarray}
where  $f_{\sigma}(x',y')$ and $g_{\sigma}(x',y')$ are given by Eqs. (\ref{f}) and (\ref{g}), respectively.
 We define
\begin{eqnarray}
X_{mp}&=&\frac{k_{mp}w}{2}\sin\theta\cos\phi,\\
Y_{mp}&=&\frac{k_{mp}\ell}{2}\sin\theta\sin\phi,\\
X_{mp,\sigma}&=&\frac{m\pi}{2}+\sigma X_{mp},\\
Y_{mp,\sigma}&=&\frac{p\pi}{2}+\sigma Y_{mp}.
\end{eqnarray}
Then, the $x$ and $y$ components of the magnetic surface current density can be easily evaluated. Aside from the usual overall constants, in the radiation zone, the contribution to the electric vector potential from the $(m,p)$th mode is
\begin{eqnarray}
{\bm F}_{mp}({\bm x},t)&\rightarrow&-\frac{\epsilon_0\tilde{v}}{16\pi r}\tilde{E}_{0mp}e^{i(k_{mp}r-n\omega_Jt)}\nonumber\\
& &\times S_{mp}^M(\theta)(\hat{\bm x}M_{mp}^x+\hat{\bm y}M_{mp}^y),\\
M_{mp}^x&=&i\sum_{\sigma,\sigma'=\pm}\sigma' e^{iY_{mp}\sigma}e^{i\sigma'm\pi x_m/w}\nonumber\\
& &\times\sin\Bigl(\frac{p\pi}{2}+\frac{\sigma p\pi y_p}{\ell}\Bigr)\nonumber\\
& &\times\Bigl(\frac{\sin X_{mp,\sigma'}}{X_{mp,\sigma'}}\Bigr),\\
M_{mp}^y&=&i\sum_{\sigma,\sigma'=\pm}\sigma' e^{i\sigma X_{mp}}e^{i\sigma' p\pi y_p/\ell}\nonumber\\
& &\times\sin\Bigl(\frac{m\pi}{2}+\frac{\sigma x_m\pi m}{w}\Bigr)\nonumber\\
& &\times\Bigl(\frac{\sin Y_{mp,\sigma'}}{Y_{mp,\sigma'}}\Bigr).
\end{eqnarray}
We first preserve the von Neumann boundary condition, so that we average $dP/d\Omega$ in the Model I procedure over both $x_m$ and $y_p$:  $\langle dP(x_m)/d\Omega\rangle_{\rm I}=\frac{1}{2}[dP(0)/d\Omega+dP(w/m)/d\Omega]$ for $m$ odd and $\langle dP(x_m)/d\Omega\rangle_{\rm I}=\frac{1}{2}[dP(-w/2m)/d\Omega+dP(w/2m)/d\Omega]$ for $m$ even.  Similarly,  $\langle dP(y_p)/d\Omega\rangle_{\rm I}=\frac{1}{2}[dP(0)/d\Omega+dP(\ell/p)/d\Omega]$ for $p$ odd and $\langle dP(y_p)/d\Omega\rangle_{\rm I}=\frac{1}{2}[dP(-\ell/2p)/d\Omega+dP(\ell/2p)/d\Omega]$ for $p$ even.  In Model II, we average over all allowed values of $-w/m\le x_m\le w/m$ and $-\ell/p\le y_p\le\ell/p$, and the results are given in the Appendix.

\begin{figure}
\includegraphics[width=0.45\textwidth]{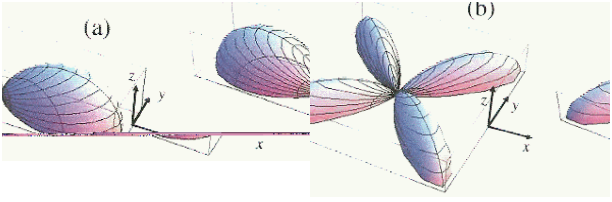}
\caption{(color online) Spherical plots of the electromagnetic cavity radiation from a square mesa with $\ell/w=1$ and $n_r=4.2$  suspended in vacuum. (a) The TM$^z_{2,0}$ mode  (b) The TM$^z_{2,2}$ mode.}\label{fig6}
\end{figure}

Three-dimensional plots of the radiation from the TM$^z_{2,0}$ mode and  the degenerate TM$^z_{0,8}$ mode for $\ell/w=4$ and $n_r=4.2$ for a sample suspended in vacuum as calculated in Model I are shown in Fig. 3.  Spherical plots of the TM$^z_{2,1}$ and TM$^z_{2,2}$ modes with $\ell/w=4$ and $n_r=4.2$ are shown in Fig. 4.  We note that the radiation from the TM$^z_{2,0}$ mode has a maximum in the $xz$ plane, whereas the radiation from the degenerate TM$^z_{0,8}$ mode vanishes in the $xz$ plane and has a maximum at a finite $\theta$ value in the $yz$ plane.  For the TM$^z_{2,1}$ and TM$^z_{2,2}$ modes with slightly higher energies, the radiation patterns shown in Fig. 4 indicate that there is also no radiation in either the $xz$ and $yz$ planes.

In addition, in anticipation that angular measurements of the radiation from square mesas might soon be made \cite{Kadowaki3}, we plotted the anticipated radiation from the cavity excitation of a square mesa with $\ell/w=1$ and $n_r=4.2$ using Model I for a sample suspended in vacuum.  In Fig. 5(a), we plotted the radiation for the fundamental TM$^z_{1,0}$ mode. This mode has maximal radiation in the $xz$ plane. The degenerate TM$^z_{0,1}$ mode exhibits radiation that is precisely the same as in Fig. 5(a) but rotated by $90^{\circ}$ about the $z$ axis.  In Fig. 5(b), the radiation from the cavity TM$^z_{1,1}$ mode is shown.  The radiation from this mode is invariant under rotations of $90^{\circ}$ about the $z$ axis, as expected from symmetry arguments, is small in both the $xz$ and $yz$ planes, and vanishes at $\theta=0^{\circ}$.  In Fig. 6, the radiation of the TM$^z_{2,0}$ and TM$^z_{2,2}$ modes of a square mesa with $n_r=4.2$ are shown.  As seen from Fig. 6(a), the radiation from the cavity TM$^z_{2,0}$ mode  vanishes  in the $yz$  plane, and symmetry thus requires that the radiation from the cavity TM$^z_{0,2}$ mode vanishes in the $xz$ plane.  The cavity radiation from the TM$^z_{2,2}$ mode vanishes in both the $xz$ and $yz$ planes, as shown in Fig. 6(b).

\begin{figure}
\includegraphics[width=0.45\textwidth]{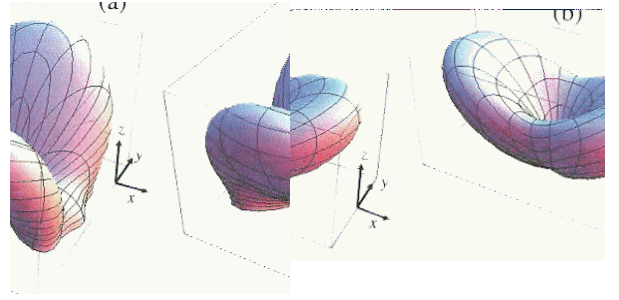}
\caption{(color online) Spherical plots of the  combined cavity radiation from the degenerate TM$^z_{2,0}$ and TM$^z_{0,8}$ cavity modes in rectangular mesas with integral $\ell/w=4$ and $n_r=4.2$. (a) $|\tilde{E}_{020}|/|\tilde{E}_{008}|$=1 for a sample suspended in vacuum. (b) $|\tilde{E}_{020}|/|\tilde{E}_{008}|$=0.1 for a sample on a superconducting substrate.}\label{fig7}
\end{figure}

\begin{figure}
\includegraphics[width=0.45\textwidth]{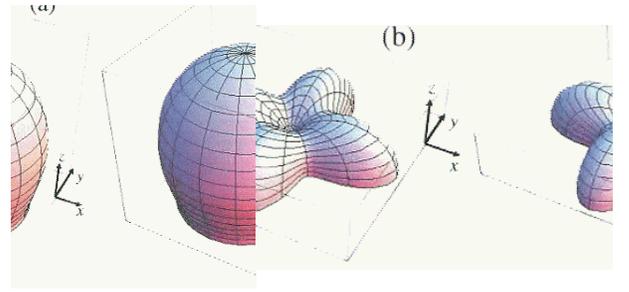}
\caption{(color online) Spherical plot of the  combined cavity radiation from the degenerate TM$^z_{n,0}$ and TM$^z_{0,n}$ cavity modes in square mesas with  $\ell/w=1$ and $n_r=4.2$ suspended in vacuum.   (a)  The combined TM$^z_{1,0}$ and TM$^z_{0,1}$ modes with $|\tilde{E}_{010}|/|\tilde{E}_{001}|=1$. (b) The combined TM$^z_{2,0}$ and TM$^z_{0,2}$ modes with $|\tilde{E}_{020}|/|\tilde{E}_{002}|=1$.}\label{fig8}
\end{figure}

\begin{figure}
\includegraphics[width=0.45\textwidth]{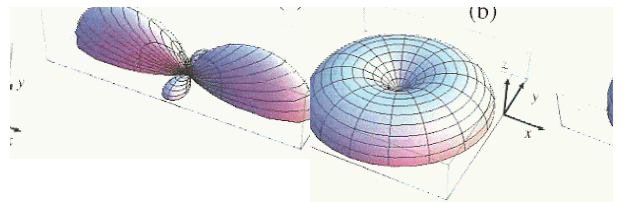}
\caption{(color online) Spherical plots of the  uniform $ac$ Josephson current radiation from rectangular mesas with integral $\ell/w$ and $n_r=4.2$ suspended in vacuum. (a) $\ell/w=4$ with the frequency locked onto that of the TM$^z_{2,0}$ cavity mode.   (b) $\ell/w=1$ with the frequency locked onto that of the TM$^z_{1,0}$ cavity mode.}\label{fig9}
\end{figure}

To illustrate the effect of a combination of the radiation from the degenerate cavity modes, in Fig. 7, three-dimensional plots of the combined cavity radiation from the TM$^z_{2,0}$ and TM$^z_{0,8}$ modes of a sample  with $\ell/w=4$ and $n_r=4.2$  are shown.  In Fig. 7(a),  $|\tilde{E}_{008}|/|\tilde{E}_{020}|=1$ for a sample suspended in vacuum.  For an equal weight of the two contributions, the largest radiation occurs for a narrow range of $\theta$ values in the $yz$ plane.  In Fig. 7(b), the analogous combined cavity radiation with $|\tilde{E}_{008}|/|\tilde{E}_{020}|=0.1$ for a sample on a superconducting substrate is pictured.  In Fig. 8, three-dimensional plots of the equally combined radiation from the TM$^z_{n,0}$ and TM$^z_{0,n}$ modes with $n_r=4.2$ for a square sample suspended in vacuum are pictured. In Fig. 8(a), the combined TM$^z_{1,0}$ and TM$^z_{0,1}$ cavity mode radiation with $|\tilde{E}_{010}|/|\tilde{E}_{001}|=1$ is shown.  In this case, there is no azimuthal anisotropy, and the axial anisotropy is rather weak.  The radiation is a maximum at $\theta=0^{\circ}$, as for cylindrical cavities in the cylindrical TM$^z_{1,1}$ mode \cite{KK2,Kadowaki3}. In Fig. 8(b), the combined radiation from the cavity TM$^z_{2,0}$ and TM$^z_{0,2}$ modes with  $|\tilde{E}_{002}|/|\tilde{E}_{020}|=1$ for the square sample suspended in vacuum  is also shown.  For these modes, the radiation exhibits $C_{4v}$ point group  symmetry, but is highly anisotropic, not only in the axial direction, but also azimuthally.

However, as it has been established that the radiation from existing mesas of Bi-2212 arises from a dual-source mechanism, with the primary radiation source being the uniform $ac$ Josephson current \cite{Kadowaki2,Kadowaki3}, and further supported by the four fits to the existing data shown in Figs. 1 and 2, it is important to remember that the predicted electromagnetic cavity radiation displayed in  Figs. 3 - 6 for rectangular mesas with $\ell/w = 4, 1$ is only part of the predicted radiation.  In addition, there will be the radiation from the uniform $ac$ Josephson current source with the same frequency.  In Fig. 9, we  displayed the radiation patterns corresponding to  those patterns in Figs. 3 - 6.  In Fig. 9(a), the  radiation from the uniform $ac$ Josephson current source in a rectangular mesa with $\ell/w = 4$ and $n_r = 4.2$ suspended in vacuum with the frequency locked onto that of the cavity TM$^z_{2,0}$ mode is shown.  This pattern is exactly the same as that for the frequency locked onto the TM$^z_{0,8}$ mode, as those frequencies are identical.  Moreover, the patterns for the slightly different frequencies locked onto the TM$^z_{2,1}$ and TM$^z_{2,2}$ modes are almost indistinguishable from that shown in Fig. 9(a), and hence are not shown.  Thus, although the radiation patterns from the excitation of TM$^z_{2,p}$  electromagnetic cavity modes with $p=0, 1, 2$ are extremely different, the patterns predicted to arise from the primary radiation source, the uniform $ac$ Josephson current, are nearly identical.  We note that for both sources, the radiation at $\theta=0^{\circ}$ vanishes, unlike the case of the electromagnetic cavity radiation for the rectangular mesa with mean $\ell/w=400/77.4$, the fits to which are displayed in Figs. 1 and 2.  In addition, we note that if the sample were placed upon a superconducting substrate, the radiation patterns would all vanish at $\theta=90^{\circ}$, as in Fig. 7(b) \cite{KK,KK2}.

In Fig. 9(b), we show the predicted radiation from the uniform $ac$ Josephson current source from a square mesa ($\ell/w=1$) with $n_r=4.2$ suspended in vacuum, with the frequency locked onto that of the TM$^z_{1,0}$ and the degenerate TM$^z_{0,1}$ mode.  Unlike the highly asymmetric corresponding electromagnetic cavity radiation shown in Fig. 5(a) which has a maximum at $\theta=0^{\circ}$ and is strongly $\phi$-dependent, this radiation is independent of the azimuthal angle $\phi$, and vanishes at $\theta=0^{\circ}$, just as for the $ac$ Josephson current source radiation, the frequency of which is locked onto that of the cylindrical cavity  TM$^z_{1,1}$ mode \cite{Kadowaki3}.  The lack of azimuthal anisotropy was also present for the combined radiation of the cavity TM$^z_{1,0}$ and TM$^z_{0,1}$ modes with $|\tilde{E}_{001}|/|\tilde{E}_{010}|=1$ pictured in Fig. 7(a), but in that case, the radiation was a maximum at $\theta=0^{\circ}$, and in this case, it vanishes at $\theta=0^{\circ}$.  Both cases thus are essentially the same as for the corresponding cases for the uniform $ac$ Josephson radiation locked onto the cylindrical cavity TM$^z_{1,1}$ mode \cite{Kadowaki3}. For the frequency locked onto that of the rectangular TM$^z_{1,1}$ mode, the radiation emitted from the uniform $ac$ Josephson current source is nearly identical to that pictured in Fig. 9(b), so it is not shown for brevity. In addition, the radiation from the $ac$ Josephson current source the frequency of which is locked onto either that of the cavity TM$^z_{2,0}$ mode or the cavity TM$^z_{2,2}$ mode is also independent of azimuthal angle $\phi$, so those figures are also not shown for brevity. At those frequencies, both radiation sources lead to a vanishing output at $\theta=0^{\circ}$, but each cavity source radiation is highly $\phi$-dependent, unlike the uniform $ac$ Josephson current source radiation.  As for the other rectangular mesas studied, if the square mesa were placed atop a superconducting substrate, both radiation sources would lead to a vanishing output at $\theta=90^{\circ}$.

 \section{V. Discussion and Summary}
 Since the publication of the Wang {\it et al.} paper \cite{Kleiner}, there has been an ongoing debate as to whether the radiation emitted from rectangular mesas originates partly from the standing wave across the width of the mesa or partly from excited  waves of the same wavelength $\lambda=2w$ that appear to form along the length of the rectangular mesa.  Since experiments on a large number of rectangular mesas have determined that the wavelength of the radiation $\lambda=2w$ \cite{Ozyuzer,Kadowaki,Kadowaki2}, it is natural to suppose that the cavity mode to which the $ac$ Josephson current couples is the TM$^z_{1,0}$ mode.  However, laser measurements have suggested that the standing waves forming along the length of the mesa appear to have the same wavelength $\lambda\approx 2w$.  Since the mesas in which these standing waves have been detected have since been shown to emit radiation, it is important for our understanding of the radiation mechanism to determine which, if either, of these possibilities is correct.  We therefore re-examined the published angular dependence of the radiation emitted from a rectangular mesa \cite{Kadowaki2}.  Although one might suppose that it would be easy to distinguish these models from the calculated output radiation, for the dual-source mechanism with an arbitrary mix of radiation from the uniform $ac$ Josephson current source and the radiation from the excited electromagnetic cavity source, we found that both scenarios are equally probable:  it is impossible from the existing data to distinguish whether the cavity source forms with the wave across the width or along the length of the rectangular mesa.  This conclusion is based upon two models of the coherent cavity and $ac$ Josephson radiation sources contributing incoherently to the output, as required for any reasonable fit to experiment, since the data for positive and negative $\theta$ values were nearly symmetric both in the $xz$ and $yz$ planes \cite{Kadowaki2}.
The equally probable overall best fits to the data of Kadowaki {\it et al.} from a mesa with mean $\ell/w=400/77.4$ are for the standing wave forming  across the mesa width in Model I   and for the traveling wave forming along the sample length in Model II. The first case is consistent with the analysis of Kadowaki {\it et al.}\cite{Kadowaki2}, preserving the von Neumann boundary conditions.  In the second case, the traveling wave is free to move along the length of the sample with no preferred nodal position.

In addition, there is a physical difference in the relative importance of the cavity portion of the overall output radiation of the two radiation sources.  If the excited cavity waves were to form across the sample width, about half of the total radiation would arise from each source.  But if the excited cavity waves were to form along the sample length, about two-thirds of the radiation would arise from the $ac$ Josephson current source.

We remark that there are many ways to do the symmetric averaging of the cavity mode radiation in the presence of the radiation from the uniform $ac$ Josephson current source.  In this work, we only used Models I and II.  For Model I with the wave along the length, we assumed the wave was centered either symmetrically, for $\ell/w$ slightly larger than an even integer, or asymmetrically, for $\ell/w$ slightly larger than an odd integer, as for the sample under study.  Then, we assumed $\langle dP_{1,\ell}(y_n)/d\Omega\rangle_{\rm I}=\frac{1}{2}[dP(0)/d\Omega+dP(w/n)/d\Omega]$ for $n$ odd and $\frac{1}{2}[dP_{1,\ell}(-w/2n)/d\Omega+dP_{1,\ell}(w/2n)/d\Omega]$ for $n$ even.  This results in a contribution to the output from the two-source mechanism that is symmetric about $\theta=0^{\circ}$, as in experiment \cite{Kadowaki2}.  It also preserves the von Neumann boundary condition.  We note, however, that this Model I fit for a standing wave along the length led to the worst fit of the four models, but the Model II fit for a traveling wave along the length led to a tie for the best overall fit to the data.  Hence, some other averaging of $y_1$ that preserves the even $\theta$ symmetry of the combined output of the two-source mechanism could lead to an improved overall best fit.  For example, one could in principle treat the combined equal outputs from $y_1$ and $y_1+w$, as in the experiment, treating $y_1$ as a  fitting parameter in addition to ${\cal A}_{1,\ell}$ and ${\cal B}_{1,\ell}$, and perform a three-parameter fit to the data.  Such a three-parameter fit would likely lead to a lower standard deviation than that for the Model II fit, in which $y_1$ was averaged over all allowed values.  This optimum fixed value of $y_1$ would lead to a standing wave along the mesa length. However,  only about 1/3 of the combined output arose from the excited cavity mode in the Models I and II fits.  Hence, modifying the fit with a third adjustable parameter that improves the minor output fraction without making some comparable adjustment to the major (uniform $ac$ Josephson current source) output fraction could be unreliable. A more reliable three-parameter fit might therefore arise from the inclusion of a non-uniform correction to the $ac$ Josephson current source \cite{KK2}.  Hence the question as to whether the optimum fit for a wave along the rectangle length would be standing or traveling is unresolved.

 We then considered the possibility of investigating mesas with different $\ell/w$ ratios.  We found that for $n\ell/w=2q$, where $q$ is any integer and $n$ is the index of the TM$^z_{n,0}$ rectangular cavity mode, the electromagnetic cavity radiation from a standing wave along the length of the mesa (in the degenerate TM$^z_{0,2q}$ mode)  vanishes in the $xz$ plane.    In those cases, any observed radiation in the $xz$ plane would have to arise from either the TM$^z_{n,0}$ electromagnetic cavity mode or the uniform $ac$ Josephson radiation.  In addition, for $n$ even, the radiation from the cavity TM$^z_{n,0}$ mode with a wave across the width vanishes in the $yz$ plane. We  presented figures  of the predicted radiation for a rectangular sample with $\ell/w=4$ and for a square sample, both  suspended in vacuum, to illustrate these points.  Since for integral $\ell/w$, the radiation can occur from either or both of the two degenerate cavity modes, we pictured examples of the combined cavity radiation from both modes, which has reduced azimuthal anisotropy, especially when $\ell/w=1$.  We strongly encourage further experiments on samples with even-integral $\ell/w$ ratios to confirm these predictions.  It would be very interesting to compare samples with even-integral $\ell/w$ values with those having slightly different $\ell/w$ values, to see if the radiation pattern is substantially different.

The authors thank Krsto Ivanovic, Reinhold Kleiner, and Masashi Tachiki for helpful discussions.  This work has been supported in part by CREST-JST (Japan Science and Technology Agency), WPI (World Premier International Research Center Initiative)-MANA (Materials Nanoarchitectonics) project (NIMS) and Strategic Initiative category (A) at the University of Tsukuba.

 \section{Appendix}
 First, we list the results of our calculations for the radiation from a general cavity mode for an arbitrary $\ell/w$ ratio.
 For Model I, we have
\begin{eqnarray}
C_{n,i}^{\rm I}&=&A^2_{n,i},\label{Cni}\\
 D_{n,i}^{\rm I}&=&B^2_{n,i},\label{Dni}\\
 E^{\rm I}_{n,i}&=&2A_{n,i}B_{n,i}\label{Eni}
 \end{eqnarray}
  for $i=w,\ell$, where
 \begin{eqnarray}
 A_{n,w}&=&\sin Y_n\sum_{\sigma=\pm}\frac{\sigma^n\sin(X_{n,\sigma})}{X_{n,\sigma}},\label{Anw}\\
 B_{n,w}&=&2\frac{\sin Y_n}{Y_n}\sin\Bigl(\frac{n\pi}{2}+X_n\Bigr),\label{Bnw}\\
A_{n,\ell}&=&2\frac{\sin X_n}{X_n}\left\{\begin{array}{cc}
 \sin Y_n\cos\Bigl(\frac{n\pi\ell}{2w}\Bigr)&\hskip10pt n\>{\rm even}\\
 \cos Y_n\sin\Bigl(\frac{n\pi\ell}{2w}\Bigr)&\hskip10pt n\>{\rm odd}\end{array}\right.,\label{Anl}\\
 B_{n,\ell}&=&\sin X_n\sum_{\sigma=\pm}\frac{\sigma^n\sin Y_{n,\sigma}}{Y_{n,\sigma}}.\label{Bnl}
 \end{eqnarray}
 For Model II,
 \begin{eqnarray}
 C_{n,w}^{\rm II}&=&\sin^2Y_n\sum_{\sigma=\pm}\frac{\sin^2X_{n,\sigma}}{X^2_{n,\sigma}},\label{CnwII}\\
 D_{n,w}^{\rm II}&=&\frac{\sin^2Y_n}{Y^2_n}\sum_{\sigma=\pm}\sin^2X_{n,\sigma},\label{DNwII}\\
 E_{n,w}^{\rm II}&=&2\frac{\sin^2Y_n}{Y_n}\sum_{\sigma=\pm}\frac{\sigma\sin^2X_{n,\sigma}}{X_{n,\sigma}},\label{EnwII}\\
 C_{n,\ell}^{\rm II}&=&\frac{\sin^2X_n}{X^2_n}\sum_{\sigma=\pm}\sin^2Y_{n,\sigma},\label{CnlII}\\
 D_{n,\ell}^{\rm II}&=&\sin^2X_n\sum_{\sigma=\pm}\frac{\sin^2Y_{n,\sigma}}{Y^2_{n,\sigma}},\label{DnlII}\\
  E_{n,\ell}^{\rm II}&=&2\frac{\sin^2X_n}{X_n}\sum_{\sigma=\pm}\frac{\sigma\sin^2Y_{n,\sigma}}{Y_{n,\sigma}}.\label{EnlII}
 \end{eqnarray}

 Then, we list the results for the radiation from the TM$^z_{mp}$ mode for a rectangle of length $\ell$ and width $w$.
 Then, in Model I, we have
\begin{eqnarray}
C_{mp}^{\rm I}=\langle|M^x_{mp}|^2\rangle_{\rm I}&=&\sin^2\Bigl(\frac{p\pi}{2}+Y_{mp}\Bigr)\nonumber\\
& &\times\Bigl(\sum_{\sigma=\pm}\frac{\sigma^m\sin X_{mp,\sigma}}{X_{mp,\sigma}}\Bigr)^2,\nonumber\\
& &\\
D_{mp}^{\rm I}=\langle|M^y_{mp}|^2\rangle_{\rm I}&=&\sin^2\Bigl(\frac{m\pi}{2}+X_{mp}\Bigr)\nonumber\\
& &\times\Bigl(\sum_{\sigma=\pm}\frac{\sigma^p\sin Y_{mp,\sigma}}{Y_{mp,\sigma}}\Bigr)^2,\nonumber\\
& &\\
E_{mp}^{\rm I}=\langle M^x_{mp}M^{y*}_{mp}+c.c.\rangle_{\rm I}&=&2\Bigl(\sum_{\sigma=\pm}\frac{\sigma\sin^2X_{mp,\sigma}}{X_{mp,\sigma}}\Bigr)\nonumber\\
& &\times\Bigl(\sum_{\sigma'=\pm}\frac{\sigma'\sin^2Y_{mp,\sigma'}}{Y_{mp,\sigma'}}\Bigr).\nonumber\\
\end{eqnarray}
In Model II, we find
\begin{eqnarray}
C_{mp}^{\rm II}&=&2\sin^2\Bigl(\frac{p\pi}{2}+Y_{mp}\Bigr)\nonumber\\
& &\times\Bigl(\sum_{\sigma=\pm}\frac{\sin X_{mp,\sigma}}{X_{mp,\sigma}}\Bigr)^2,\\
D_{mp}^{\rm II}&=&2\sin^2\Bigl(\frac{m\pi}{2}+X_{mp}\Bigr)\nonumber\\
& &\times\Bigl(\sum_{\sigma=\pm}\frac{\sin Y_{mp,\sigma}}{Y_{mp,\sigma}}\Bigr)^2,\\
E_{mp}^{\rm II}&=&2\Bigl(\sum_{\sigma=\pm}\frac{\sigma\sin^2X_{mp,\sigma}}{X_{mp,\sigma}}\Bigr)\nonumber\\
& &\times\Bigl(\sum_{\sigma'=\pm}\frac{\sigma'\sin^2Y_{mp,\sigma'}}{Y_{mp,\sigma'}}\Bigr).
\end{eqnarray}
Then, the differential power per unit solid angle emitted from the TM$^z_{mp}$ cavity mode is proportional to
\begin{eqnarray}
\frac{dP_{mp}^j}{d\Omega}&\propto& |S^M_{mp}(\theta)|^2\Bigl(C_{mp}^j+D_{mp}^j-\sin^2\theta[C_{mp}^j\cos^2\phi\nonumber\\
& &+D_{mp}^j\sin^2\phi-E_{mp}^j\sin\phi\cos\phi]\Bigr)
\end{eqnarray}
for $j$ = I, II.

\end{document}